\documentclass[journal=jpcbfk,manuscript=article,layout=twocolumn]{achemso}
\setkeys{acs}{articletitle=true}
\usepackage[T1]{fontenc}
\usepackage{lmodern}
\usepackage{graphicx}
\usepackage{amsmath}
\usepackage{booktabs}
\usepackage{siunitx}
\usepackage{url}
\usepackage{longtable}
\newif\iffullsi
\fullsitrue
\graphicspath{{figures/}}

\author{Lucia Mal\'{\i}\v{c}kov\'{a}}
\email{lucia.malickova@modelosinteligenciaartificial.es}
\affiliation{Modelos Inteligencia Artificial, C/ Tajinaste, 54, Pta.~G4, 38620 San Miguel de Abona (Santa Cruz de Tenerife), Spain}

\title{Countering Anion Inhibition of Laccase Electron Transfer: Electrostatic Design, Charge-Transfer
Dynamics on a Superconducting Quantum Processor, and Its Limits}

\begin{document}

\begin{abstract}
Hydroxide at neutral-to-alkaline pH and halides inhibit fungal laccases and limit their use in wastewater treatment.
From GFN2-xTB cluster models of \textit{Trametes versicolor} laccase (PDB 1GYC) we build a four-site Hubbard model of
the His454--T3A--T3B--His452 junction of the trinuclear copper cluster and execute its charge-transfer dynamics on the
Euro-Q-Exa quantum computer (IQM, 8 qubits, Trotterized circuits with Pauli twirling; 726 jobs). The charge-transfer
frequency is measured to \SIrange{3}{5}{meV}, reproducibly over nine days and across changes of the device qubit map:
hydroxide raises the charge-transfer energy of the crystal-geometry model by \SI{92.0 \pm 2.2}{meV} (exact 87) and exceeds fluoride by
\SI{10.1 \pm 1.8}{meV}. A screen of charged probes in three dielectric models identifies Y116E; in the probe model the
hardware resolves a reduction of the hydroxide penalty by \SI{74 \pm 7}{meV} and of the halide penalty by
\SI{64 \pm 5}{meV}, acting on the site-energy gap rather than on the coupling. A real Glu116 (five conformers) with
PROPKA protonation states and the coupled Asp456 gives a realistic reduction of \SIrange{10}{30}{meV}, with the same
sign in 28 of 28 model combinations; the direction and ordering are the prediction, rate factors only an illustration. Scanning the on-site repulsion reverses the sign of the hydroxide effect, and a
one-electron Hartree model reproduces the reversal: the measured dynamics is effectively single-particle. On a
Fermi--Hubbard benchmark the hardware does not resolve electron correlation; it would need a contrast loss below
0.13\% per CZ gate instead of the measured 0.34--0.54\%.
\end{abstract}

\section{Introduction}

Laccases (EC 1.10.3.2) are multicopper oxidases that couple the one-electron oxidation of phenols, anilines and
related compounds at a type-1 (T1) copper to the four-electron reduction of O$_2$ to water at a trinuclear copper
cluster (TNC) formed by one type-2 (T2) and two type-3 (T3) coppers.\cite{Solomon1996,Solomon2014,Jones2015}
Electrons reach the TNC from T1 through a conserved Cys--His pathway of about \SI{13}{\angstrom}: the T1 ligand
Cys453 is flanked in sequence by His452 and His454, which coordinate the two T3 coppers.\cite{Jones2015,Piontek2002}
Fungal laccases with a high T1 potential, such as the enzyme from the white-rot fungus \textit{Trametes versicolor},
are attractive green oxidants for pulp bleaching, lignin processing and the removal of dyes, phenolic pollutants,
pharmaceuticals and other micropollutants from municipal and industrial wastewater.\cite{Rodgers2010,Gasser2014,Adamian2021}

Two inhibitions limit such applications. First, the activity of fungal laccases on phenolic substrates follows a
bell-shaped pH profile whose alkaline side is attributed to hydroxide binding at the TNC, which slows internal
electron transfer from T1.\cite{Xu1997} Second, halides inhibit the enzymes, fluoride most strongly, then chloride and
bromide, an order that has been linked to access of the anion to the T2/T3 site.\cite{Xu1996,Valles2020} Kinetic work on
\textit{T. versicolor} laccase found that chloride and hydroxide act through a common mechanism and that chloride
inhibition is strongest at low pH (competitive $K_i$ = 0.35~mM at pH~3 and 23.7~mM at pH~6).\cite{Raseda2014} Both
effects are relevant for water treatment, where the enzymes have to operate near neutral pH and in the presence of
salts.\cite{Gasser2014,Valles2020} Mutations near the copper sites shift the pH optimum of fungal and
bacterial laccases and their halide sensitivity,\cite{Xu1998,Kolyadenko2023,Pardo2015} and QM/MM studies have
explored how mutations near T1 change its redox potential,\cite{Gotze2016} which makes the electrostatic environment of
the TNC a target for rational design.

Electron transfer (ET) in proteins is described by Marcus theory, in which the rate depends on the electronic
coupling between donor and acceptor, the driving force and the reorganization energy.\cite{Marcus1985} The coupling is
mediated by the protein through specific pathways,\cite{Beratan1992,Onuchic1992} and the distance dependence of
biological ET rates is well established.\cite{Moser1992,Winkler2014} The same quantities, site energies and couplings,
define a tight-binding (Hubbard) model of the junction between the His bridges and the T3 coppers.\cite{Hubbard1963}
The real-time dynamics of such a model can be simulated on a digital quantum computer,\cite{Lloyd1996,McArdle2020,Bauer2020}
and metalloenzymes are prominent targets of quantum computation.\cite{Reiher2017} Recent hardware studies have treated
iron--sulfur clusters with a quantum processor coupled to a supercomputer,\cite{IBMRiken2025} the conformers of a
miniprotein\cite{Trpcage2025} and small molecules on IQM hardware.\cite{IQMchem2026} At the same time, the evidence for
a generic exponential quantum advantage in quantum chemistry is weak,\cite{Lee2023} and present processors operate
before fault tolerance with error mitigation.\cite{Kim2023,Cai2023} It is therefore important to state precisely what a
present-day quantum processor contributes to a biochemical question.

In this work we use the Euro-Q-Exa quantum computer to address anion inhibition of laccase ET in three steps.
(i) We execute the charge-transfer (CT) dynamics of an enzyme-derived four-site model of the laccase TNC junction on
the hardware and show that the CT energy, and its change by hydroxide and halides, are measured at meV precision,
reproducibly over days and across reconfigurations of the device. (ii) We screen electrostatic modifications of the
second coordination sphere and identify a mutation, Y116E, that removes most of the CT penalty imposed by hydroxide
and halides; we test it on the hardware, decompose its mechanism, and follow it from a generic charge probe to a
real glutamate with explicit conformers and protonation states. (iii) We measure where the hardware stops resolving
electron correlation, using the junction model itself and a Fermi--Hubbard benchmark. We state explicitly that the
four-site dynamics is classically exact and, as we show, effectively single-particle; the hardware results validate
the execution of the model and quantify the limits of the device, while the design conclusions rest on the classical
model and its stated approximations.

\section{Methods}

\subsection{Cluster models}
Cluster models were cut from the 1.90-\AA{} crystal structure of \textit{T. versicolor} laccase with a full complement
of coppers (PDB 1GYC).\cite{Piontek2002} The pathway cluster contains the T1 copper with its His395 and His458
imidazoles and Cys453, the His452--Cys453--His454 tripeptide with its backbone, the three TNC coppers with their His
imidazoles, the bridging hydroxide of the T3 pair and the T2-bound ligand (H$_2$O, OH$^-$, F$^-$ or Cl$^-$). The
protein-embedded models add the charged second-sphere residues closest to the T2 water, Asp77 (\SI{5.2}{\angstrom}),
Asp424 (\SI{5.5}{\angstrom}) and Arg243 (\SI{7.6}{\angstrom}), as capped side chains; with the probes and Asp456 the
clusters contain 124--141 atoms. The cluster is a quartet (Cu(I) T1 and three Cu(II)) with the T3 pair antiparallel
(broken-symmetry). Hydrogen positions were relaxed with GFN2-xTB\cite{Bannwarth2019} while heavy atoms were kept at the
crystal positions.

\subsection{Tight-binding parameters}
Site energies and couplings were obtained from the self-consistent GFN2-xTB Fock matrix $F$ and the overlap $S$ in
the minimal valence basis. After L\"owdin orthogonalization,\cite{Lowdin1950} $F' = S^{-1/2} F S^{-1/2}$, the block of
$F'$ belonging to each pathway fragment was diagonalized and one frontier orbital $c_i$ selected per site: for Cu sites
the fragment orbital with more than 30\% Cu~3d weight closest to the Fermi level, for the bridges (Cys453, His452,
His454) the highest occupied fragment orbital. Site energies and couplings are $\varepsilon_i = c_i^\dagger F' c_i$ and
$t_{ij} = c_i^\dagger F' c_j$. Because each site orbital comes from a separate diagonalization, the sign of each hopping
is arbitrary; populations, currents and frequencies are gauge invariant on a chain, so the observables used here do
not depend on it. Dielectric screening by the protein and solvent was modeled with the analytical linearized
Poisson--Boltzmann model ALPB\cite{Ehlert2021} for a protein-like environment ($\varepsilon \approx 4$, parametrized
as diethyl ether) and for water, and compared with the gas phase. Because the effective screening at this site is not known, we bracket it by the gas phase (no screening) and water
(full screening) and use $\varepsilon \approx 4$ as a protein-like middle value. The size, and in some cases the sign,
of second-sphere effects depends on screening, so we accepted a design conclusion only if it had the same sign in all
three dielectric models.

\subsection{Mutants, probes and protonation}
Second-sphere mutants were modeled explicitly (D77N, D424N, R243A, D77N/D424N; Q70K, Q102K and Q70K/Q102K, with
CH$_3$NH$_3^+$ at the Gln CD/OE1 position). To screen positions rather than specific side chains, generic charge probes
(CH$_3$COO$^-$ or CH$_3$NH$_3^+$ with the charged group at the tip of the native side chain) were placed at Tyr116,
Leu459 and Phe450, the residues closest to His454 and T3A that a charged side chain could replace. Real Glu116 was then
built on the crystal backbone (N, C$_\alpha$, C$_\beta$ kept) on a \ang{10}/\ang{10}/\ang{15} grid of
$\chi_1/\chi_2/\chi_3$; conformers with heavy-atom contacts below \SI{3.0}{\angstrom} (\SI{2.55}{\angstrom} for a
carboxylate O to an N/O partner) were rejected, crystal waters being assumed displaced. The 317 clash-free conformers
all lie in the region of the Tyr116 side chain ($\chi_1$ = \ang{30}--\ang{150}; Tyr116 has \ang{93}); five conformer
families, with the carboxylate \SIrange{5.3}{7.4}{\angstrom} from T3A, were represented by their least strained member,
and two of them were also modeled as protonated Glu. p$K_a$ values were estimated with PROPKA~3\cite{Olsson2011} on the
full protein with Glu116 inserted, and Asp456, which couples to Glu116, was added to the cluster in both protonation
states. Anion affinities relative to the wild type (WT) are
$\Delta\Delta E = [E(\mathrm{mut},X) - E(\mathrm{mut},\mathrm{H_2O})] - [E(\mathrm{WT},X) - E(\mathrm{WT},\mathrm{H_2O})]$
from GFN2-xTB single points in the three dielectric models; $\Delta\Delta E > 0$ means weaker binding of $X$ in the
mutant. These are differences of electronic energies at fixed geometries, without relaxation, entropy or explicit water;
we use only their sign, not their magnitude.

\subsection{Junction model}
\begin{figure}
\includegraphics[width=\columnwidth]{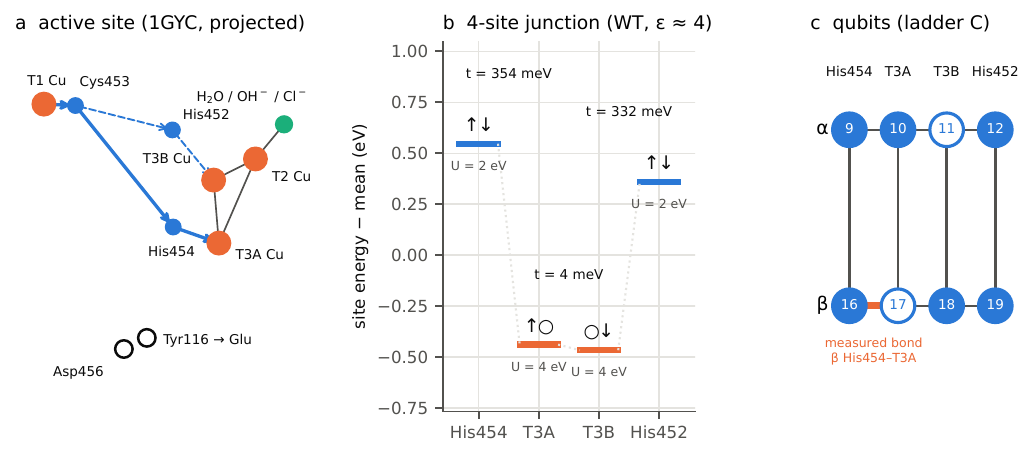}
\caption{Model. (a) Active site of \textit{T. versicolor} laccase (PDB 1GYC), crystal coordinates projected on their principal
plane: Cu ions, Cys453 S, His452/His454 ring centroids, the T2 ligand and the Tyr116/Asp456 positions; arrows: the
T1 $\to$ Cys--His $\to$ T3 pathway. (b) Four-site junction (WT, $\varepsilon \approx 4$) with GFN2-xTB site energies
(relative to their mean), couplings $t$ and on-site repulsions $U$; arrows: occupied spin orbitals, open circles:
holes. (c) Mapping of the $\alpha$ and $\beta$ chains onto the qubits of ladder C of EQE1; rungs carry the on-site
interaction; the measured $\beta$ His454--T3A bond is highlighted.}
\label{fig:system}
\end{figure}

The junction His454--T3A--T3B--His452 is described by a four-site Hubbard Hamiltonian\cite{Hubbard1963}
\begin{equation}
H = \sum_{i,\sigma} \varepsilon_i n_{i\sigma} + \sum_{\langle ij\rangle,\sigma} t_{ij}
(c^\dagger_{i\sigma} c_{j\sigma} + \mathrm{h.c.}) + \sum_i U_i n_{i\uparrow} n_{i\downarrow},
\end{equation}
with nearest-neighbour hopping from the fragment couplings and on-site repulsion $U$ = 4~eV on Cu and 2~eV on the His
bridges. The Cu value lies in the range of effective one-band repulsions obtained for Cu~3d holes from first principles
and spectroscopy (about 3.7--5~eV; the bare 3d repulsion of three-band models is larger, 7--10~eV);\cite{Sheshadri2022}
the bridge value is a chosen smaller value for a delocalized $\pi$ orbital. Both are model choices; the $U$ scan below
shows that the sign of the anion effect holds for any common scaling of $U$ above 0.55 of these values. The initial state has both His bridges doubly occupied and one hole per spin on the T3 coppers (the
antiferromagnetic T3 pair: $\beta$ hole on T3A, $\alpha$ hole on T3B; Figure~\ref{fig:system}). Starting the pair as
singlet, triplet ($M_S = 0$) or broken-symmetry product gives identical charge dynamics to below \SI{e-3}{e} within
3~fs, because the direct T3A--T3B coupling (\SIrange{4}{8}{meV}) is too weak to matter on this time scale; extending
the model by T2 and T1 adds no femtosecond dynamics either (couplings to the core $\leq$ 0.03~eV). The
Jordan--Wigner transformation\cite{Jordan1928} maps the $\alpha$ and $\beta$ chains onto the two rows of a qubit
ladder (8 qubits); the rungs carry the on-site interaction. For the classical ET analysis the same parameters were used
in a larger pathway model (T1, Cys453, His454, His452, T3A, T3B) to obtain the T1$\to$T3 driving force, an effective
donor--acceptor coupling by L\"owdin partitioning and Marcus rates for reorganization energies $\lambda$ = 0.5--1.1~eV.

\subsection{Circuits}
Time evolution uses second-order Strang splitting\cite{Strang1968} of $H$ into even bonds, odd bonds and on-site
terms, with a step of 1~fs. Each hopping term is a native XX+YY rotation and each on-site term a ZZ rotation with
single-qubit phases, both compiled to two CZ gates on neighbouring qubits of a ladder; the deepest circuits (6~fs)
contain 128 CZ gates. Circuits were built with Qiskit\cite{Qiskit2024} and verified against exact dynamics computed
with ffsim. Ladders were selected from calibration data (readout and relaxation errors per qubit) on the square lattice
of the device; most runs used two ladders, $\alpha$ 38--41/$\beta$ 44--47 (``B'') and $\alpha$ 9--12/$\beta$ 16--19
(``C''), in the numbering of the 52-qubit map. When a qubit was lost or restored and the numbering changed, circuits
were re-indexed so that they ran on the same physical qubits.

\subsection{Observables and error mitigation}
Three observables were used. (1) Site populations in the computational basis. (2) Spin-resolved bond currents and
bond coherences, obtained with a number-conserving XX+YY($\pi/2$, $\beta$) rotation on a bond before measurement:
$\beta = 0$ gives $\langle n_i - n_j\rangle = 2\,\mathrm{Im}\langle c_i^\dagger c_j\rangle$ (current) and
$\beta = \pi/2$ gives $-2\,\mathrm{Re}\langle c_i^\dagger c_j\rangle$ (coherence). (3) The CT frequency: the complex
bond signal $\rho(t) = \langle c^\dagger_{\mathrm{His454}} c_{\mathrm{T3A}}\rangle_\beta$ is reconstructed from the two
bases at 24 times between 0.25 and 6~fs and fitted jointly to
$\mathrm{Im}\,\rho = a \sin(\omega t) e^{-\gamma t}$, $\mathrm{Re}\,\rho = b (1 - \cos \omega t) e^{-\gamma t} + c$.
$\hbar\omega$ is the CT excitation energy of the junction model, a model quantity without a direct spectroscopic
counterpart whose changes, not its absolute value, are interpreted; $\gamma$ absorbs the hardware damping. Hardware noise
damps the amplitude but hardly shifts the frequency, which is why the frequency became the main observable. The same
fit applied to the exact signal and to the noiseless circuits on the same time grid gives the reference values; the
difference between the two references is the Trotter error. The penalty of an anion $X$ is
$\hbar\omega(X) - \hbar\omega(\mathrm{H_2O})$; variants compared in a result were always measured in the same jobs.

Every CZ gate was Pauli-twirled\cite{Wallman2016} (12 random instances per circuit), which turns coherent errors into
stochastic ones common to all variants. Results were post-selected on the conserved particle numbers
$(N_\alpha, N_\beta)$; the fraction of shots that survives is a model-free indicator of circuit quality. In early runs
linear zero-noise extrapolation (ZNE) by gate folding\cite{Temme2017} was also tested. Errors are bootstrap estimates
over twirl instances,\cite{Efron1979} which include shot noise and the spread between instances; results from several
runs were combined with inverse-variance weights and errors scaled by $\sqrt{\chi^2/\mathrm{dof}}$ where larger than
one.

\subsection{Population decomposition}
The frequency gives only $\Omega = \sqrt{\Delta^2 + 4t^2}$, where $\Delta$ is the effective gap between the His and Cu
sites and $t$ the effective coupling. The population of the T3A hole oscillates with amplitude $A = 4t^2/\Omega^2$,
so amplitude and frequency together give $t$ and $\Delta$ separately, the two Marcus ingredients (rate $\propto t^2$,
driving force from $\Delta$). These runs used the same circuits measured in the Z basis, with $t = 0$ circuits to
correct preparation and readout errors.

\subsection{Hardware}
Euro-Q-Exa is a EuroHPC quantum computer hosted by the Leibniz Supercomputing Centre and integrated with the
SuperMUC-NG supercomputer.\cite{EuroQExa} System~1 (backend EQE1) has 54 physical superconducting qubits from IQM on a
square lattice with tunable couplers\cite{IQM2024} and was accessed through the Munich Quantum Software
Stack.\cite{MQSS2025} Depending on the calibration, 51--53 qubits were available during the project. Between 19
October and 12 November 2026 we ran 63 experiments with 726 jobs, 68\,300 circuits and about $1.03 \times 10^8$ shots
(Table~S1).

\subsection{Correlation benchmark}
To test the device on genuinely correlated dynamics, we used a one-dimensional Fermi--Hubbard chain with $t = 0.3$~eV
(the scale of the enzyme couplings), $U = 0$ or $4t$, and $N = 4$ or 6 sites (8 or 12 qubits) on nested windows of the
longest straight ladder of the device, starting from a charge domain wall (left half doubly occupied). The observable
is the charge imbalance $(N_\mathrm{left} - N_\mathrm{right})/N$ after post-selection, compared with exact dynamics and
with time-dependent mean field (Hartree).\cite{GoogleHubbard2020} A second $N = 4$ window on other qubits separates
the effect of size from that of qubit quality.

\section{Results}

\subsection{Hydroxide slows the T1$\to$T3 step in the classical model}
In the pathway model, OH$^-$ on T2 lowers the T1$\to$T3 driving force from $-0.365$ to $-0.300$~eV and slows the
Marcus rate by a factor of 1.9--2.9 for reorganization energies between 0.5 and 1.1~eV
($2.9 \times 10^{10}$ vs $1.1 \times 10^{10}$~s$^{-1}$ at $\lambda = 0.8$~eV), consistent with the activity drop of
fungal laccases at neutral and alkaline pH.\cite{Xu1997} F$^-$ and Cl$^-$ give the same rate as OH$^-$
($1.1 \times 10^{10}$~s$^{-1}$): at the level of the ET step, the anion's charge, not its chemistry, sets the effect.
On the femtosecond scale the junction shows the same trend: at 2~fs the T3 copper has gained 0.444~e from His454
with H$_2$O and 0.325~e with OH$^-$. The absolute rates are faster than experiment, as expected for semiempirical
couplings; the relative effect of the ligand is the result.

\subsection{Junction dynamics on the hardware}
The first hardware runs measured site populations of the junction (Figure~S6). With H$_2$O the hardware followed the
exact curve within 0.03~e up to 2~fs (68 CZ, 31--36\% post-selection survival), but the raw data reproduced the pH
effect (H$_2$O $-$ OH$^-$) poorly ($\chi^2$ = 147 for 4 points), because coherent errors differed between the two
variants. Pauli twirling made the errors stochastic and common to both, and with twirling and ZNE the pH effect was
reproduced at all times ($\chi^2$ = 1.7, $p$ = 0.79), including its rise and decay up to 3.5~fs (Figure~S7). The
predicted sign change at 4~fs (128 CZ) was not reproduced: at that depth the spin density on His454 no longer refocused, which defined a certified window
of 0--3.5~fs ($\leq$ 128 CZ) for population observables.

Bond currents proved more robust. The $\beta$ current His454$\to$T3A agreed with the exact model over 0.5--3~fs
($\chi^2$ = 13.3 for 12 points), with a peak at 1~fs of \SI{0.34 \pm 0.03}{e/fs} for H$_2$O (exact 0.38) and
\SI{0.29 \pm 0.02}{e/fs} for OH$^-$ (exact 0.31); the population change obtained by integrating the measured current
satisfied the continuity equation ($\chi^2$ = 13.8/12). The currents expected to vanish ($\alpha$ His454--T3A and
$\beta$ T3B--His452) averaged \SI{0.003}{e/fs}, against about \SI{0.3}{e/fs} in the allowed channels (Figure~S8): only
the $\beta$ electron can enter T3A through His454 and only the $\alpha$ electron T3B through His452. This selection
rule follows from the occupations of the initial state and, as discussed below, is also the reason why the measured
dynamics is effectively single-particle. Running three copies of the junction simultaneously on disjoint ladders
(24 qubits) confirmed that the selection rule is independent of the layout, but showed that post-selection survival
and damping differ strongly between ladders (41/27/17\%), and that linear ZNE no longer restores the signal below
about 30\% survival (Figure~S9).

\subsection{Charge-transfer energy and its hydroxide shift}
\begin{figure}
\includegraphics[width=\columnwidth]{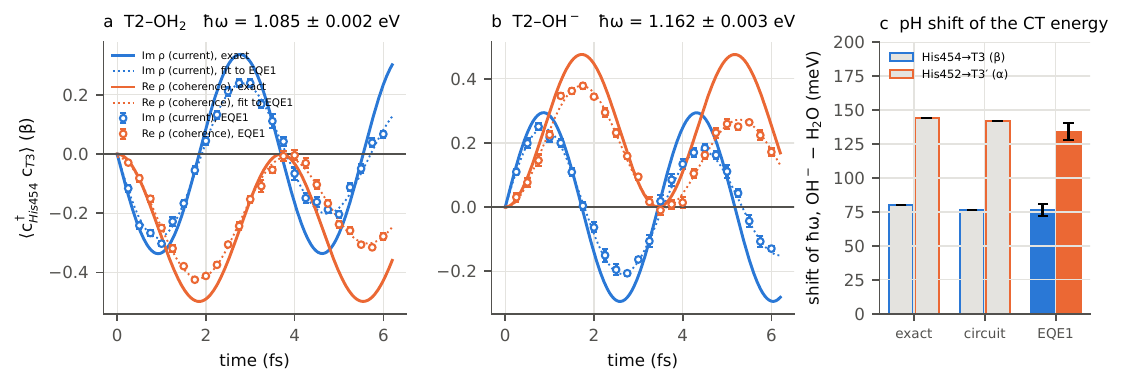}
\caption{Charge-transfer signal of the His454--T3A bond on EQE1 (bare pathway model, no protein field; ladder A, 0.25--6~fs,
raw twirled data). Points: hardware imaginary and real parts of $\rho(t) = \langle c^\dagger_{\mathrm{His454}}
c_{\mathrm{T3A}}\rangle_\beta$ with bootstrap errors; solid lines: exact dynamics; dotted lines: damped-oscillation
fits. The fitted CT energies $\hbar\omega$ for H$_2$O and OH$^-$ at T2 and their difference (the OH$^-$ penalty) are
compared with the exact model and the noiseless circuits.}
\label{fig:freq}
\end{figure}

The CT frequency was the most precise observable (Figure~\ref{fig:freq}). On the bare pathway model, the hardware
frequencies agreed with the noiseless circuits to 0.2--0.3\%. Over six runs on three ladders, hydroxide raised the CT
energy of the His454$\to$T3A channel by \SI{74.8 \pm 2.2}{meV} and that of the His452$\to$T3B channel by
\SI{138 \pm 6}{meV} (noiseless circuits 76 and 142, exact 80 and 144~meV). The two channels therefore respond
differently to the same ligand, with an asymmetry of \SI{58 \pm 7}{meV} (circuits 66~meV). The remaining 3\%
difference between the circuits and the exact dynamics is the Trotter error of the 1-fs step, not hardware error.

\begin{figure}
\includegraphics[width=\columnwidth]{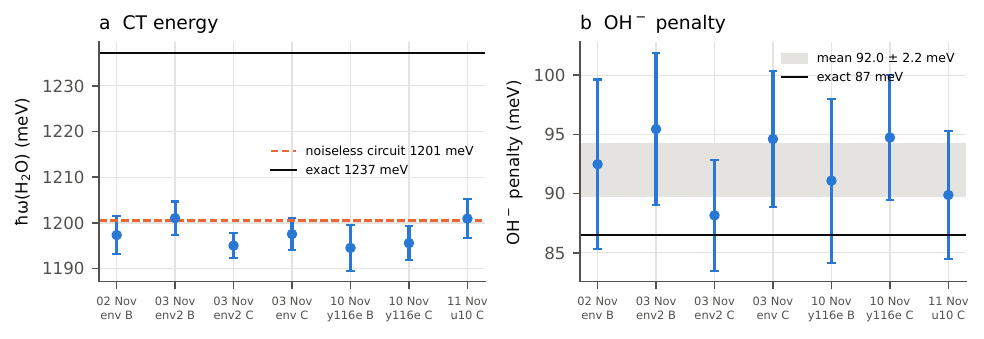}
\caption{Stability of the reference measurement (protein-embedded WT model, $\varepsilon \approx 4$). Seven independent
submissions on ladders B and C between 2 and 11 November 2026, spanning three changes of the device qubit map.
(a) CT energy with H$_2$O; dashed: noiseless circuit, solid: exact model. (b) OH$^-$ penalty; grey band: weighted
mean $\pm$ error. Error bars: bootstrap over twirl instances.}
\label{fig:stability}
\end{figure}

With the protein environment (Asp77, Asp424, Arg243, $\varepsilon \approx 4$) the model penalty of the His454 channel
is 87~meV (circuit 83). This configuration served as the reference of all later comparisons and was measured in seven
independent submissions on two ladders between 2 and 11 November 2026, across three changes of the qubit map
(Figure~\ref{fig:stability}). The CT energy with H$_2$O was \SI{1197.2 \pm 1.4}{meV} (scatter \SI{2.7}{meV}; circuit
1200, exact 1237~meV) and the OH$^-$ penalty \SI{92.0 \pm 2.2}{meV} (scatter \SI{2.8}{meV}, $\chi^2/\mathrm{dof}$ =
0.27). The scatter is below the statistical errors: no drift is detectable over nine days. The damping $\gamma$ varied
between 0.13 and 0.24~fs$^{-1}$ without affecting the frequency. Across all runs the per-ladder systematic floor of a
single frequency difference is about \SIrange{3}{5}{meV}, and \SIrange{5}{10}{meV} for differences that involve
different jobs.

\subsection{Anion fingerprint}
\begin{figure}
\includegraphics[width=\columnwidth]{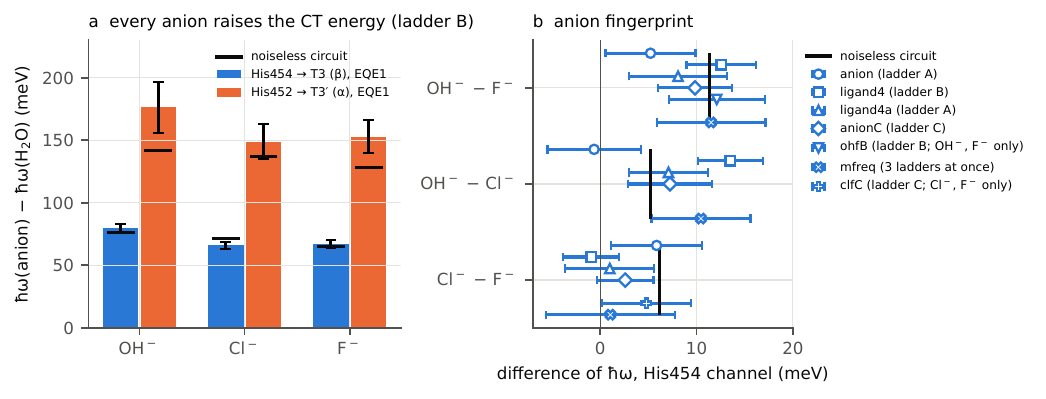}
\caption{Anion fingerprint (bare pathway model). (a) CT energy change relative to H$_2$O for OH$^-$, Cl$^-$ and F$^-$ on
one ladder, with noiseless-circuit values (horizontal bars). (b) Differences between anions from all runs and ladders
(points; run names as in Table S1) with their combined value; OH$^-$ is separated from both halides, Cl$^-$ and F$^-$ are not separated.}
\label{fig:fingerprint}
\end{figure}

Every anion raised the CT energy (Figure~\ref{fig:fingerprint}). Combined over six runs on three ladders, OH$^-$ exceeded
F$^-$ by \SI{10.1 \pm 1.8}{meV} (5.5$\sigma$; noiseless circuit 11~meV) and Cl$^-$ by \SI{8.4 \pm 2.3}{meV}, whereas
Cl$^-$ and F$^-$ were indistinguishable (\SI{1.9 \pm 1.6}{meV}; circuit 6~meV). The charge of the T2 ligand thus sets
about 90\% of the CT penalty and its chemical identity the remaining \SI{10}{meV}. Because F$^-$ and Cl$^-$ act alike
on the CT step, the strong difference between their inhibition constants\cite{Xu1996,Raseda2014} must arise from
binding and access to the TNC rather than from the ET step itself.

\subsection{Second-sphere mutants: positive charges}
Removing the negative charges near T2 (D77N, D424N, D77N/D424N) lowered the OH$^-$ penalty at $\varepsilon \approx 4$
but raised it in water, so no mutant was proposed from these. Two lysines replacing Gln70 and Gln102 (tips 7.8 and
\SI{8.1}{\angstrom} from the T2 water) lowered the penalty in all three dielectric models (exact 87$\to$71~meV at
$\varepsilon \approx 4$). On the hardware, four runs per system on two ladders gave \SI{92.0 \pm 2.9}{meV} for WT and
\SI{70.8 \pm 2.9}{meV} for Q70K/Q102K, a reduction of \SI{21.2 \pm 4.1}{meV} (5.1$\sigma$; exact 16, circuit 13~meV);
the halide penalty dropped by \SI{13.9 \pm 4.0}{meV} (exact 14~meV). The single mutants each lowered the penalty
(Q70K \SI{24.0 \pm 5.6}{meV}, Q102K \SI{17.3 \pm 4.9}{meV}; exact 9 and 7~meV), and the ordering of their CT
energies with H$_2$O (Q102K $<$ Q70K $<$ double) was reproduced, but their additivity was not resolved at the
systematic floor. However, the change in anion affinity of the double mutant depends on the dielectric model
($\Delta\Delta E$ for OH$^-$ = $-62$, $-15$ and $+1.4$~kcal/mol in gas, $\varepsilon \approx 4$ and water): in a
protein-like environment the lysines would bind the inhibiting anions more strongly. Lowering the CT penalty is
therefore not sufficient; a useful mutation must also not strengthen anion binding.

\subsection{A negative charge near His454: the probe screen}
\begin{figure*}
\includegraphics[width=\textwidth]{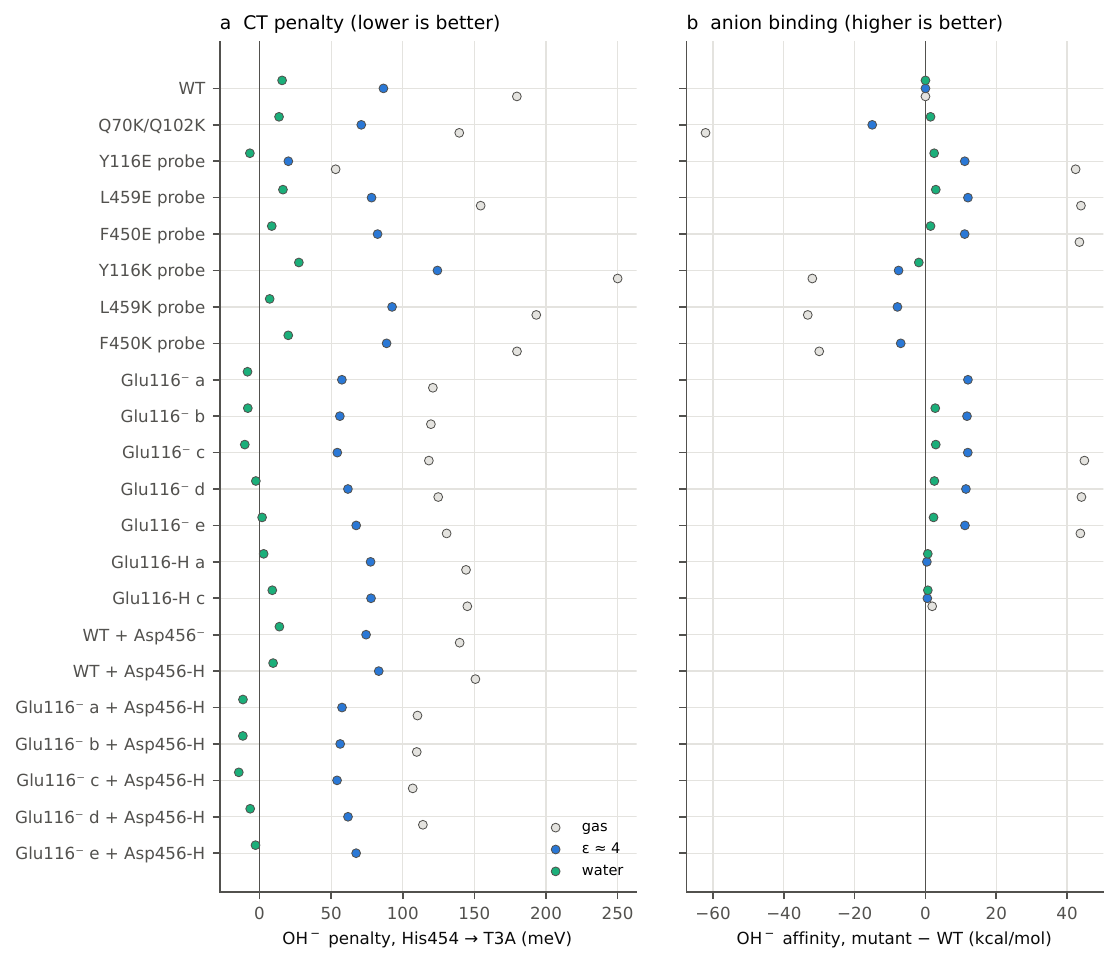}
\caption{Classical screen of second-sphere modifications (GFN2-xTB clusters; gas, $\varepsilon \approx 4$, water).
(a) OH$^-$ penalty on the His454 channel (exact four-site dynamics); lower is better. (b) OH$^-$ affinity relative to
WT, $\Delta\Delta E$ (electronic energies, sign only); higher means weaker binding. Rows: WT; positive charges
(Q70K/Q102K, K probes); negative probes; real Glu116 conformers a--e; protonated Glu116; WT and Glu116 with Asp456
in both protonation states. Missing points: SCF not converged or affinity not computed.}
\label{fig:screen}
\end{figure*}

We therefore required two conditions in all three dielectric models: a lower OH$^-$ penalty on the His454 channel and
weaker OH$^-$ binding (Figure~\ref{fig:screen}). Positive probes (Y116K, L459K, F450K) increased anion binding in every
model. Negative probes at Leu459 and Phe450 weakened OH$^-$ binding (+11 to +12~kcal/mol at $\varepsilon \approx 4$) but
hardly changed the penalty (78 and 82 vs 87~meV). Only the probe at Tyr116 satisfied both conditions: penalty
53/20/$-7$~meV (gas/$\varepsilon \approx 4$/water; WT 180/87/16~meV) and OH$^-$ binding weaker by +42/+11/+2.5~kcal/mol.
Tyr116 lies next to His454 and T3A (its hydroxyl \SI{5.9}{\angstrom} from T3A), on the side of the junction opposite to
the T2 ligand. The halide penalties of the Y116E probe model were also close to zero at $\varepsilon \approx 4$
($-6$ and $-5$~meV for Cl$^-$ and F$^-$, WT 56 and 53~meV).

\subsection{Y116E on the hardware}
\begin{figure*}
\includegraphics[width=\textwidth]{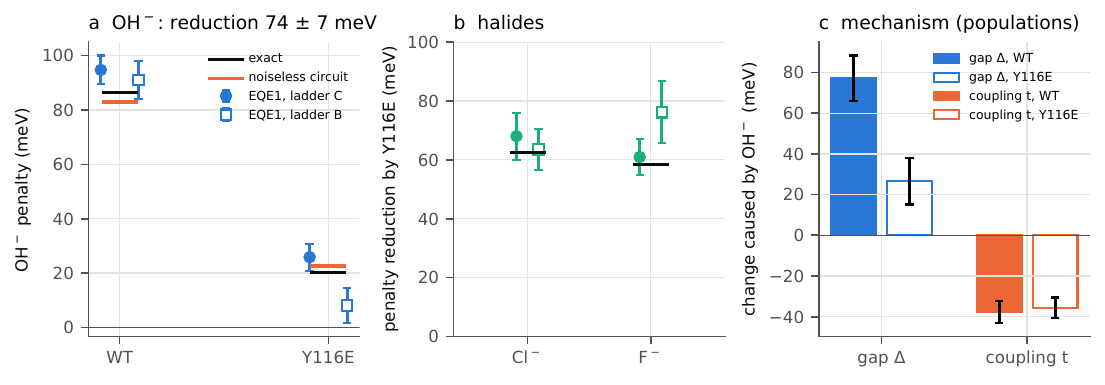}
\caption{Y116E (probe model, $\varepsilon \approx 4$) on EQE1, WT and mutant in the same jobs. (a) OH$^-$ penalty on
ladders C (filled) and B (open); horizontal bars: exact (black) and noiseless circuit (orange). (b) Reduction of the
Cl$^-$ and F$^-$ penalty by Y116E on both ladders, relative to the H$_2$O difference of panel a; black bars: exact.
(c) Change of the gap $\Delta$ and of the coupling $t$ caused by OH$^-$, from Z-basis populations combined over both
ladders: Y116E removes most of the gap increase and leaves the coupling change unchanged.}
\label{fig:y116e}
\end{figure*}

With WT and Y116E in the same jobs on two ladders, the OH$^-$ penalty fell from \SI{93 \pm 4}{meV} to
\SI{19 \pm 9}{meV}, a reduction of \SI{74 \pm 7}{meV} (exact 66, circuit 60~meV); the two ladders agreed within
1.2$\sigma$ (Figure~\ref{fig:y116e}a). The CT energy with H$_2$O itself dropped by \SIrange{94}{101}{meV} (exact 116,
circuit 106~meV), showing that the carboxylate stabilizes the whole His454$\to$T3 manifold. For the halides, the
difference $\hbar\omega(X)_\mathrm{WT} - \hbar\omega(X)_\mathrm{Y116E}$ was \SI{163.1 \pm 3.0}{meV} for Cl$^-$ and
F$^-$ together, and \SI{98.7 \pm 4.2}{meV} for H$_2$O, so Y116E lowered the halide penalty by \SI{64 \pm 5}{meV}
(Cl$^-$ 65 $\pm$ 6, F$^-$ 64 $\pm$ 7; exact 62 and 58, circuit 57 and 53~meV; Figure~\ref{fig:y116e}b). Cl$^-$ and F$^-$ were again
indistinguishable, both in WT (\SI{-2 \pm 4}{meV}) and in the mutant (\SI{-2 \pm 7}{meV}). Within the probe model at crystal geometry, Y116E removes
essentially the whole CT penalty of all three anions; the relaxation tests below show that this overstates the effect.

\subsection{Mechanism: the gap, not the coupling}
The population runs separate the gap $\Delta$ and the coupling $t$ (Figure~\ref{fig:y116e}c). In WT, OH$^-$ raised
$\Delta$ by \SI{77 \pm 11}{meV} and lowered $t$ by \SI{38 \pm 6}{meV}; in Y116E, OH$^-$ raised $\Delta$ by only
\SI{27 \pm 12}{meV} and lowered $t$ by \SI{36 \pm 5}{meV}. The mutant thus removed \SI{51 \pm 16}{meV} of the
OH$^-$-induced gap increase (3.2$\sigma$; circuit 57~meV) and left the coupling change untouched
($-3 \pm 8$~meV). Both ladders gave the same decomposition. Hydroxide inhibits the CT step mainly by raising the site
energy that the hole has to overcome, and the carboxylate counteracts exactly this quantity; the mechanism is
electrostatic and does not involve the electronic coupling through the His bridge. The OH$^-$-induced decrease of
$t$ is about twice the circuit value on both ladders, which may reflect a systematic loss of amplitude on the
hardware (amplitudes about 0.6--0.7 of the circuit values) and is not interpreted further.

\subsection{From the probe to a real glutamate}
The probe places a full negative charge where the Tyr116 hydroxyl is; a real Glu116 is shorter. For the five conformer
families of Glu116$^-$, the OH$^-$ penalty at $\varepsilon \approx 4$ was 54--68~meV, a reduction of
\SIrange{19}{32}{meV} instead of 66~meV, while OH$^-$ binding remained weaker by +11.2 to +12.0~kcal/mol in every family
(Figure~\ref{fig:screen}; Table~S2). The conformers differ by only $\pm$7~meV, so the result does not hinge on a
particular side-chain geometry. PROPKA places the p$K_a$ of Glu116 between 5.0 and 8.9 depending on the conformer and
shows that Glu116 and Asp456 share roughly one proton: in the three families closest to T3A, Glu116 is charged above
pH~$\approx$~6.5 and raises the p$K_a$ of Asp456 from 7.2 to 8.3--9.2. A protonated Glu116 would leave only a
\SI{9}{meV} reduction and no change in binding. With Asp456 included in the cluster and the protonation states that
PROPKA assigns at pH 8--9 to the three families closest to T3A (WT: Asp456$^-$; Y116E: Glu116$^-$ with Asp456 protonated), Y116E lowers the OH$^-$ penalty by \SIrange{17}{20}{meV} at $\varepsilon \approx 4$, \SIrange{29}{33}{meV} in the gas phase and \SIrange{26}{28}{meV} in water; including the two remaining families in the same protonation state widens these ranges to \SIrange{7}{20}{meV}, \SIrange{26}{33}{meV} and \SIrange{17}{28}{meV}, and with both acids charged the reduction is \SIrange{17}{25}{meV} at $\varepsilon \approx 4$. The sign of the effect is the same for every conformer, both Asp456 states and all three dielectric models (28 of 28 combinations; the gas-phase SCF of conformer e with Asp456 did not converge). Interestingly,
the native Asp456 already lowers the WT penalty by 12~meV when charged. The realistic prediction for the mutant is
therefore a reduction of the CT penalty by \SIrange{10}{30}{meV}, about one third of the unrelaxed probe value. Partial relaxation tests this further (GFN2-xTB/ALPB at $\varepsilon \approx 4$; T2 ligand, the residue-116 group and
all hydrogens mobile, the remaining heavy atoms fixed). Hydroxide stays bound to T2 (Cu--O 2.35 $\to$ \SI{1.93}{\angstrom})
but moves towards T3A (5.9 $\to$ \SI{4.0}{\angstrom}), which more than doubles the absolute penalty (WT with Asp456$^-$:
74 $\to$ 201~meV). The reduction by the real mutant, however, is stable: \SI{20}{meV} unrelaxed and \SI{24}{meV}
relaxed (Glu116 conformer c with protonated Asp456). The probe's larger unrelaxed reduction (66~meV) shrinks to the same
\SI{24}{meV} after relaxation. Absolute penalties are therefore geometry dependent, whereas the mutant effect, the
quantity on which the design rests, is not. The real-glutamate models were not run on the hardware: the dynamics is classically exact, and the hardware
had reproduced the model differences of the probe to within \SIrange{3}{5}{meV}, so a repetition would add no
information about the mutant; the design conclusion rests on the classical model.

\subsection{Role of the on-site repulsion}
\begin{figure}
\includegraphics[width=\columnwidth]{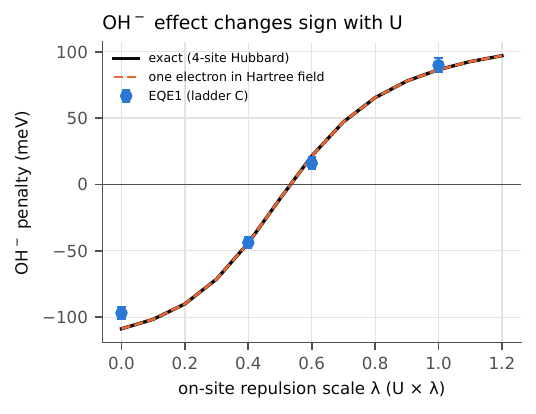}
\caption{OH$^-$ penalty versus a common scaling $\lambda_U$ of all on-site repulsions (WT, protein-embedded model, ladder
C). Points: EQE1 (both ligands in the same jobs at each $\lambda_U$); solid line: exact four-site Hubbard dynamics;
dashed line: one $\beta$ electron in the static Hartree field of the frozen $\alpha$ electrons. The two curves
coincide, so the sign change is a mean-field (charging-energy) effect, not a correlation effect.}
\label{fig:uscan}
\end{figure}

To ask whether the anion effect depends on the on-site repulsion, all $U$ were scaled by $\lambda_U$ and the WT
penalty was measured at four values on the same ladder, with both ligands in the same jobs at each value
(Figure~\ref{fig:uscan}). The sign of the OH$^-$ effect reverses: $-97 \pm 5$~meV at $\lambda_U = 0$ (21$\sigma$
below zero), $-44 \pm 4$ at 0.4, $+16 \pm 4$ at 0.6 and $+90 \pm 5$~meV at the physical value, with the crossover at
$\lambda_U = 0.55 \pm 0.01$ (exact 0.54). Without repulsion, hydroxide would accelerate the His454$\to$T3A transfer;
at the repulsion appropriate for Cu(II) it slows it, and the physical $U$ lies well inside the inhibiting regime (the
sign would reverse only for a Cu repulsion below about 2.2~eV). A one-electron model, in which the $\beta$ electron moves
in the static Hartree field $U_i n_{i\alpha}$ of the frozen $\alpha$ electrons, reproduces the exact curve point for
point. The reason is the occupation pattern: both $\alpha$ electrons of His454 and T3A are present and cannot move,
and the T3 halves are coupled by only a few meV, so within the 6-fs window the measured dynamics is that of a single
electron in a static field. The reversal therefore reflects the charging energy of the Cu site relative to the bridge,
which a purely electrostatic picture without the site's own repulsion would miss, but it is not an effect of electron
correlation. The same argument applies to all junction results above: the hardware executed a single-particle
problem, which a classical computer solves exactly.

\subsection{Hardware limits for correlated dynamics}
\begin{figure*}
\includegraphics[width=\textwidth]{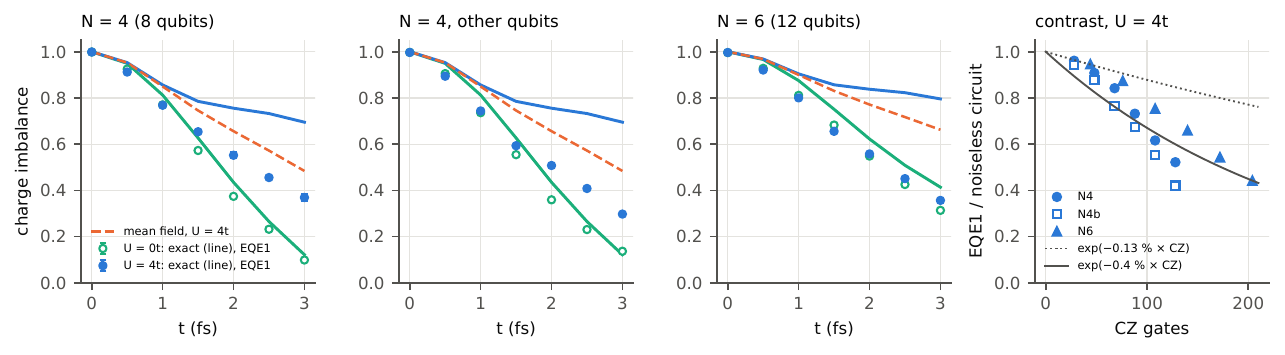}
\caption{Correlation benchmark: one-dimensional Fermi--Hubbard chain ($t = 0.3$~eV) starting from a charge domain wall,
on nested windows of the longest ladder of EQE1 (51-qubit map). First three panels: charge imbalance for $U = 0$
(open, green) and $U = 4t$ (filled, blue) with exact (solid) and time-dependent mean-field (dashed) curves, for $N = 4$
on two different qubit windows and for $N = 6$. Last panel: ratio of hardware to noiseless-circuit imbalance at
$U = 4t$ versus CZ count, with exponential decays for the measured (solid) and the required (dotted) error per CZ.}
\label{fig:limits}
\end{figure*}

The correlation benchmark was designed so that hardware noise and correlation pull in opposite directions: noise drives
the charge imbalance towards zero, as mean field does, whereas the exact correlated dynamics keeps the doubly occupied
domain in place. At $U = 4t$ and 3~fs, the exact imbalance exceeds mean field by 0.21 for $N = 4$ (128 CZ) and 0.13 for
$N = 6$ (204 CZ); at $U = 0$ the two coincide. The hardware reproduced free-fermion dynamics with an rms deviation of
0.04--0.08, but at $U = 4t$ fell below even the mean-field curve (0.37 $\pm$ 0.01 and 0.36 $\pm$ 0.01 against 0.48 and
0.66 at 3~fs; Figure~\ref{fig:limits}). The correlation gap only opens after about 1.5~fs, when the hardware deviation
already exceeds it. The ratio of the hardware to the noiseless signal decays with the number of CZ gates as
$e^{-k N_\mathrm{CZ}}$ with $k$ = 0.34--0.54\% per CZ; state preparation and readout alone cost 2.1--3.2\% per qubit;
and two windows of the same size on different qubits differ about as much as $N = 4$ and $N = 6$. Resolving the
correlation gap at 3~fs would require $k \leq 0.13\%$ for $N = 4$ and $k \leq 0.04\%$ for $N = 6$, three to eight times
better than measured. The same limit appeared in earlier attempts: a five-site model including the bridging hydroxide
of the T3 pair, whose two coppers compete for the bridge electrons (10 qubits, 80--228 CZ), gave only a 2.4$\sigma$ hint
of the predicted charge redistribution at 4--12\% survival; a model coupling the transferred charge to an effective
vibrational mode ($\lambda$ = 0.43~eV) could not be resolved because gate noise heated the mode qubits more than the
vibronic excitation itself; and linear ZNE did not restore signals below about 30\% survival. Running three copies of a
circuit in parallel increased the damping on two of three ladders, so for the frequency observable one copy per job on
the best ladder was the most efficient use of the device.

\begin{table*}
\caption{Main hardware results (EQE1, raw twirled data) with exact and noiseless-circuit references; His454$\to$T3A
channel unless noted. Penalties in meV. Errors are statistical (bootstrap over twirl instances, scaled by
$\sqrt{\chi^2/\mathrm{dof}}$ where larger than one). The systematic floor estimated from run-to-run scatter is
\SIrange{3}{5}{meV} for differences measured within the same jobs (all rows except Q70K/Q102K) and \SIrange{5}{10}{meV}
for differences between jobs (Q70K/Q102K row).}
\label{tab:summary}
\begin{tabular}{p{8cm}rrr}
\toprule
quantity & EQE1 & exact & circuit \\ \midrule
OH$^-$ penalty, bare model (6 runs) & 74.8 $\pm$ 2.2 & 80 & 76 \\
OH$^-$ penalty, His452$\to$T3B & 138 $\pm$ 6 & 144 & 142 \\
OH$^-$ penalty, protein field (7 runs) & 92.0 $\pm$ 2.2 & 87 & 83 \\
OH$^-$ $-$ F$^-$ & 10.1 $\pm$ 1.8 & 11 & 11 \\
Cl$^-$ $-$ F$^-$ & 1.9 $\pm$ 1.6 & 6 & 6 \\
reduction by Q70K/Q102K & 21.2 $\pm$ 4.1 & 16 & 13 \\
reduction by Y116E (probe), OH$^-$ & 74 $\pm$ 7 & 66 & 60 \\
reduction by Y116E (probe), halides & 64 $\pm$ 5 & 60 & 55 \\
OH$^-$ effect on $\Delta$: WT / Y116E & 77 $\pm$ 11 / 27 $\pm$ 12 & -- & 116 / 59 \\
penalty at $\lambda_U = 0$ & $-97 \pm 5$ & $-109$ & $-101$ \\
crossover $\lambda_U$ & 0.55 $\pm$ 0.01 & 0.54 & -- \\
\bottomrule
\end{tabular}
\end{table*}

\section{Discussion}

\subsection{How anions inhibit the charge-transfer step}
Within the model, any anion at T2 raises the CT energy of the His$\to$T3 junction and lowers the T1$\to$T3 driving
force, and does so mainly through the site energy (gap) of the T3 copper rather than through the electronic coupling.
About 90\% of the penalty is set by the charge of the ligand; hydroxide adds about \SI{10}{meV} over the halides, and
fluoride and chloride act alike. This assigns the different inhibition strengths of F$^-$ and Cl$^-$ observed
experimentally\cite{Xu1996,Raseda2014,Valles2020} to binding and access rather than to the ET step, and it is
consistent with a common mechanism for hydroxide and chloride inhibition.\cite{Raseda2014} The two His bridges respond
differently (His452 channel about twice as strongly), so the T3 pair is not perturbed symmetrically by the T2 ligand.

\subsection{Design rule and prediction for experiment}
The screen yields a simple design rule: a negative charge on the far side of His454/T3A from the T2 ligand stabilizes
the hole on the T3 copper, lowers the gap and offsets the anion's field, and at the same time repels the anion from T2.
Positive charges closer to T2 lower the CT penalty as well, but attract the inhibiting anion, and mutations that remove
native acids near T2 are not robust against the dielectric model and may also impair proton delivery to the TNC, in
which conserved acidic residues near T2 take part.\cite{Kolyadenko2023,Jones2015} Y116E is the only candidate that passed
both criteria in every model, and the effect survives the step from a charge probe to a real glutamate with explicit
conformers, protonation states and the coupled Asp456, at a reduced size of \SIrange{10}{30}{meV}, and survives partial relaxation (\SI{24}{meV}).

What would such a reduction mean for the rate? If a gap reduction $\delta$ translated one-to-one into the T1$\to$T3
driving force, the Marcus expression would change the rate by $\exp[\delta (\Delta G + \lambda)/(2\lambda k_B T)]$;
with $\Delta G = -0.30$~eV and $\lambda$ = 0.5--1.1~eV, $\delta$ = \SIrange{10}{30}{meV} corresponds to a factor of
about 1.1--1.5, a partial recovery of the 1.9--2.9-fold slow-down that hydroxide causes in the same model. The weaker
anion binding comes on top of this and may matter more in practice, because it lowers the fraction of enzyme with an
anion bound. The prediction is testable with standard assays: the alkaline side of the pH profile with ABTS or a
phenolic substrate, and the chloride inhibition constant at pH~6, where it is 23.7~mM for the wild-type enzyme.\cite{Raseda2014,Xu1997}
The main risk is the proton-delivery network of the TNC, which a CT model does not contain: a carboxylate next to the
cluster could change O$_2$ reduction independently of the ET step. A double check of activity at acidic pH, where the
wild type is fully active, would separate the two effects.

\subsection{What the quantum processor contributed}
The hardware executed an enzyme-derived CT model with meV precision, reproducibly over nine days, on several ladders and
through changes of the device configuration, and it resolved every model difference tested, from the 75-meV hydroxide
shift to the 10-meV difference between hydroxide and fluoride and the 50-meV gap effect of Y116E. Three practical lessons
emerged. Frequencies are far more robust than amplitudes or populations: noise damps an oscillation but hardly shifts
it. Pauli twirling is essential whenever two variants are compared, because it makes coherent errors common to both.
The post-selection survival is a free and reliable indicator of the quality of a ladder on a given day.

The hardware did not compute anything that a classical computer could not. The four-site junction is classically
exact, and in this model its dynamics is effectively single-particle. On a correlated benchmark of comparable size, the
present error rates are three to eight times too high, and the requirement grows with system size because the
correlation signal shrinks while the circuit depth grows. Models that would need the quantum computer, such as the
T3 pair with its bridging hydroxide, coupling to vibrations, or the full TNC with several mobile holes, are exactly
those that could not be run. This agrees with the general assessment that quantum advantage in chemistry requires
substantially better hardware.\cite{Lee2023} The 150-qubit second system of Euro-Q-Exa, with longer coherence
times,\cite{EuroQExa} is the natural next platform for these models, and the frequency protocol with paired variants
developed here can be transferred to them directly.

\subsection{Relevance for applications}
For wastewater treatment, laccases are typically used immobilized on supports and have to operate near neutral pH and in
the presence of salts.\cite{Gasser2014,Adamian2021} A mutation that lowers the CT penalty of hydroxide and
halides and weakens their binding addresses both limitations at once. Its benefit, however, depends on the balance
between the ET step and the other steps of the catalytic cycle, which only experiments can establish.

\section{Limitations}
(1) Cluster models of 124--141 atoms with implicit solvation; the size of second-sphere effects depends on the
dielectric model, which we address by requiring the same sign in three models. (2) All tight-binding parameters come
from GFN2-xTB; a DFT check on reduced clusters could not be converged, and a hybrid-DFT (B3LYP) test on a gas-phase
fragment confirms the direction of the Glu116 effect on the His--Cu gap but not its size (Supporting Information). (3) The on-site repulsion is chosen, not computed; the $U$ scan shows that the sign of the anion effect holds for any common scaling of $U$
above 0.55 of the chosen values (Cu repulsion above about 2.2~eV). (4) One orbital per site: the T1$\to$TNC step, whose Cys--His coupling is only
\SIrange{5}{8}{meV} in this description, is not represented; Y116E lowers the T1--T3A site-energy difference of the model by
\SIrange{26}{62}{meV} but leaves it downhill in all models. (5) Mostly static structures: the screen relaxes only hydrogens; a partial relaxation of the T2 ligand and the
residue-116 group changes the absolute penalties strongly but leaves the mutant effect unchanged (20 vs 24~meV), and
the surrounding protein never responds; empirical p$K_a$ values ($\pm$1 unit), with the Cu ions treated only
approximately by PROPKA. (6) The
reorganization energy is taken from the literature range for the rate estimate. (7) No experimental validation yet.
(8) The probe overestimates the effect of the real mutant by a factor of two to four; the probe-model hardware results
validate the execution, not the size of the real effect.

\section{Conclusions}
An enzyme-derived model of the charge-transfer junction of laccase was executed on the Euro-Q-Exa quantum computer with
meV precision, and the hardware reproduced the effects of hydroxide and halides, their fingerprint and the effect of
mutations. Classical screening in three dielectric models, followed by hardware tests and a realistic glutamate model,
identifies Y116E as a candidate that lowers the CT penalty of all three anions by \SIrange{10}{30}{meV} through the site
energy of the T3 copper and weakens hydroxide binding. The measured dynamics is effectively single-particle, and on a
correlated benchmark the hardware does not yet resolve electron correlation; the error rates required for that are three
to eight times below the present ones. The work thus provides a testable mutant prediction and a quantitative,
hardware-based statement of what present quantum processors can and cannot add to the modeling of electron transfer in
metalloenzymes.

\section{Data and code availability}
Code, quantum circuits (QPY), raw hardware results, analyses and figure scripts are available in the project repository
at \url{https://github.com/lucia-malickova/quantum}. Software versions are listed in the Supporting Information.

\begin{acknowledgement}
The author gratefully acknowledges the EuroHPC Joint Undertaking for awarding access to the Euro-Q-Exa quantum
computer under project ID EHPC-QCP-2026Q01-017 and thanks the Leibniz Supercomputing Centre (LRZ) for the
provisioning and support of the quantum computing infrastructure essential to this publication.
\end{acknowledgement}

\paragraph{Use of AI tools.} An AI assistant (Claude, Anthropic) was used for data processing: writing and running
the analysis and modeling scripts, processing the hardware results, preparing the figures and drafting and editing the
text. All quantum computations were executed on Euro-Q-Exa. The author designed the study, checked the results and is
responsible for the content.

\begin{suppinfo}
\iffullsi The Supporting Information is appended after the references (Sections S1--S9, Figures and Tables S). \fi
Supporting Information: list of all hardware runs (jobs, ladders, dates), circuit depth scan, qubit maps, junction
dynamics, bond currents, three-ladder multiplexing, earlier frequency runs, practical pH/chloride estimate, vibronic
and bridge-model attempts, Glu116 conformer and p$K_a$ tables, partial relaxation, DFT checks.
\end{suppinfo}

\iffullsi\nocite{Bannwarth2019,Cai2023,Motta2023,Olsson2011,Qiskit2024,Raseda2014,Temme2017}\fi 
\bibliography{refs}

\iffullsi
\onecolumn
\setcounter{section}{0}\setcounter{figure}{0}\setcounter{table}{0}\setcounter{equation}{0}
\renewcommand{\thesection}{S\arabic{section}}
\renewcommand{\thefigure}{S\arabic{figure}}
\renewcommand{\thetable}{S\arabic{table}}
\renewcommand{\theequation}{S\arabic{equation}}
\begin{center}
{\Large\bfseries Supporting Information}\\[4pt]
{\large Countering Anion Inhibition of Laccase Electron Transfer: Electrostatic Design, Charge-Transfer Dynamics on a
Superconducting Quantum Processor, and Its Limits}\\[4pt]
Lucia Mal\'{\i}\v{c}kov\'{a}
\end{center}

\section{Hardware runs}
Table~\ref{tab:runs} lists every run on Euro-Q-Exa system~1 (EQE1) used in this work: 63 runs, 726 jobs, 68\,300
circuits and about $1.03\times10^{8}$ shots between 19 October and 12 November 2026. Job identifiers are those of the
Munich Quantum Software Stack; the ladder is given by the $\alpha$-chain qubits in the device numbering valid on the
date of the run (the map changed between 51, 52 and 53 qubits; circuits were re-indexed to the same physical
qubits).

{\scriptsize\begin{longtable}{llrlrrll}
\caption{All Euro-Q-Exa (EQE1) runs of this work, generated from the result files.}\label{tab:runs}\\
\toprule date & run & jobs & job ids & circuits & shots & $\alpha$ qubits & result \\ \midrule\endfirsthead
\caption[]{(continued)}\\
\toprule date & run & jobs & job ids & circuits & shots & $\alpha$ qubits & result \\ \midrule\endhead
2026-10-19 & calibration\_eqe1 & 2 & 423521--429254 & 6 & 12000 & -- & device calibration \\
2026-10-19 & depth\_scan\_eqe1 & 1 & 423520 & 6 & 18000 & -- & SI: depth scan \\
2026-10-19 & ghz & 1 & 423483 & 0 & 0 & -- & device test \\
2026-10-19 & quality\_eqe1 & 2 & 423521--429254 & 0 & 0 & -- & device calibration \\
2026-10-19 & t1\_lucj\_r1\_eqe1 & 1 & 423505 & 0 & 0 & -- & SI: T1 LUCJ \\
2026-10-19 & t1\_lucj\_r1\_eqe1\_hole & 1 & 423510 & 0 & 0 & -- & SI: T1 LUCJ \\
2026-10-21 & junction\_eqe1 & 1 & 425097 & 10 & 40000 & 32--40--46--50 & populations (3.2) \\
2026-10-21 & junction\_long\_eqe1 & 3 & 425102--425104 & 240 & 600000 & 32--40--46--50 & populations (3.2) \\
2026-10-21 & junction\_twirl\_eqe1 & 1 & 425099 & 96 & 67200 & -- & populations (3.2) \\
2026-10-21 & junction\_zne\_eqe1 & 1 & 425098 & 16 & 64000 & -- & populations (3.2) \\
2026-10-21 & layout\_ab\_eqe1 & 1 & 425037 & 2 & 8000 & -- & SI: layout \\
2026-10-21 & t1\_quality\_eqe1 & 1 & 425096 & 4 & 32000 & -- & device calibration \\
2026-10-22 & current\_eqe1 & 6 & 425441--425446 & 576 & 864000 & 32--40--46--50 & currents (3.2) \\
2026-10-23 & anion\_eqe1 & 18 & 426341--426372 & 1728 & 2592000 & 31--39--45--49 & fingerprint (3.4) \\
2026-10-23 & freq2\_eqe1 & 14 & 425961--425974 & 1344 & 2016000 & 38--39--40--41 & CT energy (3.3) \\
2026-10-23 & freq3\_eqe1 & 14 & 425976--425996 & 1344 & 2016000 & 31--39--45--49 & CT energy (3.3) \\
2026-10-23 & freq6\_eqe1 & 12 & 426300--426322 & 1152 & 1728000 & 31--39--45--49 & CT energy (3.3) \\
2026-10-23 & freq\_eqe1 & 14 & 425947--425960 & 1344 & 2016000 & 31--39--45--49 & CT energy (3.3) \\
2026-10-23 & ligand4\_eqe1 & 24 & 426445--426492 & 2304 & 3456000 & 38--39--40--41 & fingerprint (3.4) \\
2026-10-23 & multiplex\_eqe1 & 12 & 425927--425945 & 864 & 1296000 & -- & three ladders (3.2) \\
2026-10-28 & bridge2\_eqe1 & 3 & 428602--428608 & 240 & 480000 & 38--39--40--41--42 & bridge model (3.12) \\
2026-10-28 & bridge\_eqe1 & 3 & 428577--428582 & 240 & 480000 & 25--32--40--46--50 & bridge model (3.12) \\
2026-10-28 & ligand4a\_eqe1 & 24 & 428533--428565 & 2304 & 3456000 & 32--40--46--50 & fingerprint (3.4) \\
2026-10-29 & anionC\_eqe1 & 18 & 429373--429438 & 1728 & 2592000 & 9--10--11--12 & fingerprint (3.4) \\
2026-10-29 & freqC\_eqe1 & 12 & 429256--429364 & 1152 & 1728000 & 9--10--11--12 & CT energy (3.3) \\
2026-10-29 & ohfB\_eqe1 & 12 & 429441--429460 & 1152 & 1728000 & 38--39--40--41 & fingerprint (3.4) \\
2026-10-30 & clfC\_eqe1 & 12 & 429471--429940 & 1152 & 1728000 & 9--10--11--12 & fingerprint (3.4) \\
2026-10-30 & mainB\_eqe1 & 12 & 429963--429996 & 1152 & 1728000 & 38--39--40--41 & CT energy (3.3) \\
2026-10-30 & mfreq\_eqe1 & 8 & 430029--430052 & 768 & 1152000 & -- & parallel copies (3.12) \\
2026-10-30 & mfreq\_eqe1L0 & 8 & 430029--430052 & 768 & 1152000 & 9--10--11--12 & parallel copies (3.12) \\
2026-10-30 & mfreq\_eqe1L1 & 8 & 430029--430052 & 768 & 1152000 & 39--40--41--42 & parallel copies (3.12) \\
2026-10-30 & mfreq\_eqe1L2 & 8 & 430029--430052 & 768 & 1152000 & 26--27--28--29 & parallel copies (3.12) \\
2026-11-02 & envB\_eqe1 & 12 & 436538--436561 & 1152 & 1728000 & 38--39--40--41 & reference, stability (3.3) \\
2026-11-03 & env2B\_eqe1 & 12 & 437275--437292 & 1152 & 1728000 & 38--39--40--41 & reference, stability (3.3) \\
2026-11-03 & env2C\_eqe1 & 12 & 437365--437386 & 1152 & 1728000 & 9--10--11--12 & reference, stability (3.3) \\
2026-11-03 & envC\_eqe1 & 12 & 437135--437154 & 1152 & 1728000 & 9--10--11--12 & reference, stability (3.3) \\
2026-11-03 & kk2B\_eqe1 & 12 & 437294--437323 & 1152 & 1728000 & 38--39--40--41 & Q70K/Q102K (3.5) \\
2026-11-03 & kk2C\_eqe1 & 12 & 437387--437406 & 1152 & 1728000 & 9--10--11--12 & Q70K/Q102K (3.5) \\
2026-11-03 & kkB\_eqe1 & 12 & 437156--437176 & 1152 & 1728000 & 38--39--40--41 & Q70K/Q102K (3.5) \\
2026-11-03 & kkC\_eqe1 & 12 & 437183--437270 & 1152 & 1728000 & 9--10--11--12 & Q70K/Q102K (3.5) \\
2026-11-08 & halB\_eqe1 & 24 & 438428--439985 & 2304 & 3456000 & 38--39--40--41 & Q70K/Q102K halides (3.5) \\
2026-11-08 & halC\_eqe1 & 24 & 439986--440009 & 2304 & 3456000 & 9--10--11--12 & Q70K/Q102K halides (3.5) \\
2026-11-09 & hal2B\_eqe1 & 24 & 440567--440616 & 2304 & 3456000 & 38--39--40--41 & Q70K/Q102K halides (3.5) \\
2026-11-09 & hal2C\_eqe1 & 24 & 440618--440664 & 2304 & 3456000 & 9--10--11--12 & Q70K/Q102K halides (3.5) \\
2026-11-09 & popB\_eqe1 & 12 & 441144--441171 & 1200 & 1800000 & 38--39--40--41 & populations (3.8) \\
2026-11-09 & popC\_eqe1 & 12 & 441174--441197 & 1200 & 1800000 & 9--10--11--12 & populations (3.8) \\
2026-11-09 & singB\_eqe1 & 24 & 440945--440993 & 2304 & 3456000 & 38--39--40--41 & single mutants (3.5) \\
2026-11-09 & singC\_eqe1 & 24 & 440996--441046 & 2304 & 3456000 & 9--10--11--12 & single mutants (3.5) \\
2026-11-09 & vibB\_eqe1 & 12 & 441198--441220 & 1200 & 1800000 & -- & vibronic model (3.12) \\
2026-11-09 & vibC\_eqe1 & 12 & 441222--441242 & 1200 & 1800000 & -- & vibronic model (3.12) \\
2026-11-10 & y116eB\_eqe1 & 24 & 442087--442117 & 2304 & 3456000 & 39--40--41--42 & Y116E (3.7) \\
2026-11-10 & y116eC\_eqe1 & 24 & 442119--442166 & 2304 & 3456000 & 9--10--11--12 & Y116E (3.7) \\
2026-11-10 & y116ehalB\_eqe1 & 24 & 442238--442261 & 2304 & 3456000 & 39--40--41--42 & Y116E halides (3.7) \\
2026-11-10 & y116ehalC\_eqe1 & 24 & 442262--442308 & 2304 & 3456000 & 9--10--11--12 & Y116E halides (3.7) \\
2026-11-10 & y116epopB\_eqe1 & 12 & 442214--442225 & 1200 & 1800000 & 39--40--41--42 & Y116E mechanism (3.8) \\
2026-11-10 & y116epopC\_eqe1 & 12 & 442226--442237 & 1200 & 1800000 & 9--10--11--12 & Y116E mechanism (3.8) \\
2026-11-11 & u00C\_eqe1 & 12 & 442886--442898 & 1152 & 1728000 & 9--10--11--12 & U scan (3.10) \\
2026-11-11 & u04C\_eqe1 & 12 & 442899--442911 & 1152 & 1728000 & 9--10--11--12 & U scan (3.10) \\
2026-11-11 & u06C\_eqe1 & 12 & 442912--442928 & 1152 & 1728000 & 9--10--11--12 & U scan (3.10) \\
2026-11-11 & u10C\_eqe1 & 12 & 442930--442943 & 1152 & 1728000 & 9--10--11--12 & U scan (3.10) \\
2026-11-12 & hubN4\_eqe1 & 4 & 443595--443598 & 336 & 504000 & 47--41--33--26 & correlation benchmark (3.11) \\
2026-11-12 & hubN4b\_eqe1 & 4 & 443599--443602 & 336 & 504000 & 26--19--12--6 & correlation benchmark (3.11) \\
2026-11-12 & hubN6\_eqe1 & 4 & 443604--443607 & 336 & 504000 & 47--41--33--26--19--12 & correlation benchmark (3.11) \\
\bottomrule\end{longtable}
}

\section{Device characterization}
\begin{figure}[h]
\includegraphics[width=\textwidth]{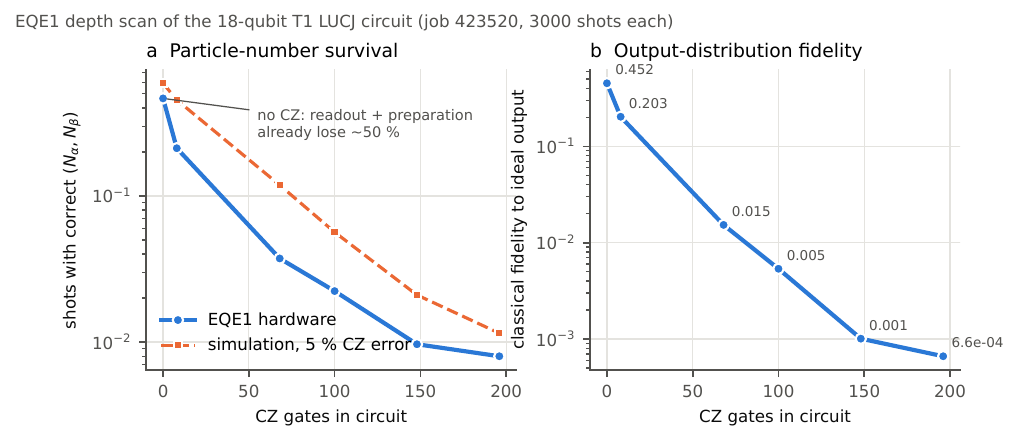}
\caption{Depth scan of an 18-qubit LUCJ circuit\cite{Motta2023} for the T1 site: particle-number survival and
output-distribution fidelity versus CZ count.}
\end{figure}
\begin{figure}[h]
\includegraphics[width=\textwidth]{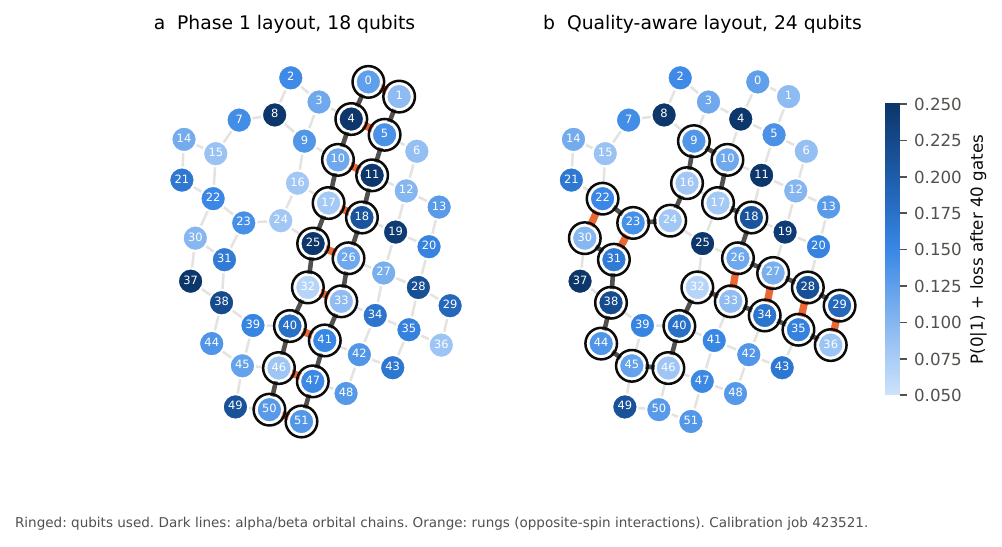}
\caption{EQE1 qubit map and the ladders used.}
\end{figure}
\begin{figure}[h]
\includegraphics[width=0.8\textwidth]{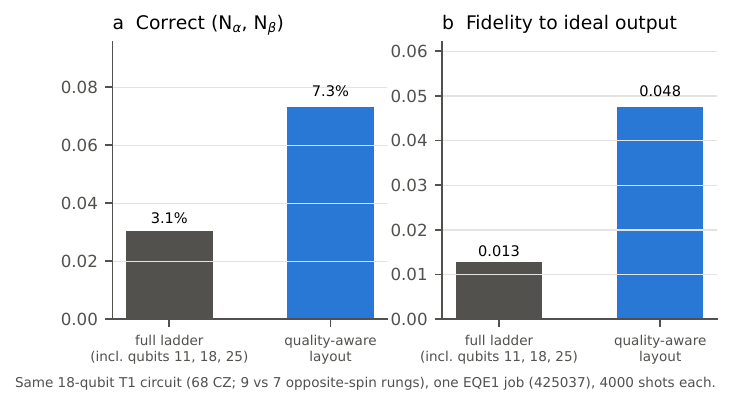}
\caption{Effect of the qubit layout (same circuit, two layouts) on the result.}
\end{figure}
\begin{figure}[h]
\includegraphics[width=0.8\textwidth]{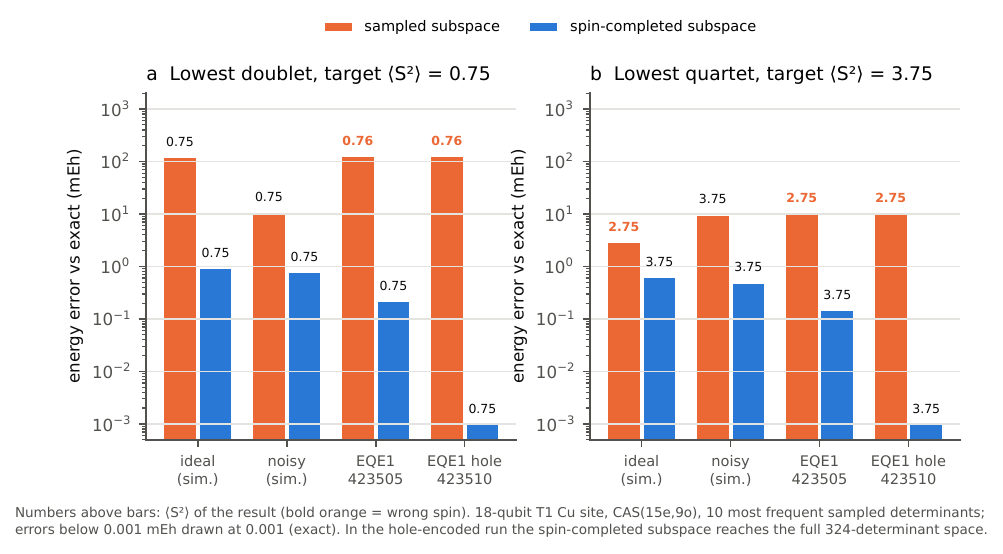}
\caption{Spin contamination of variational results for the T1 site (spin trap).}
\end{figure}
\begin{figure}[h]
\includegraphics[width=0.6\textwidth]{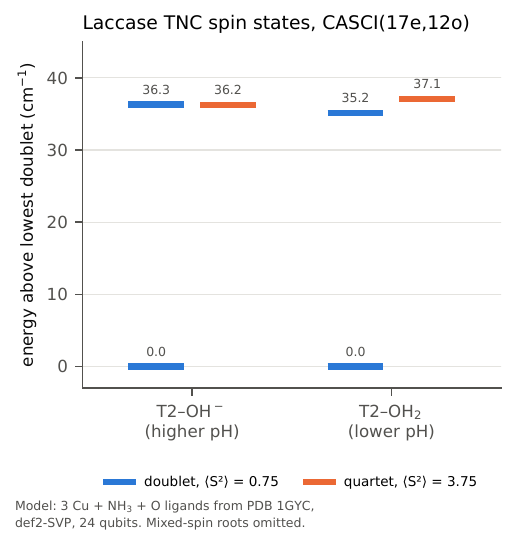}
\caption{Spin states of the TNC, CASCI(17e,12o), classical reference.}
\end{figure}

\section{Junction dynamics, currents and multiplexing}
\begin{figure}[h]
\includegraphics[width=\textwidth]{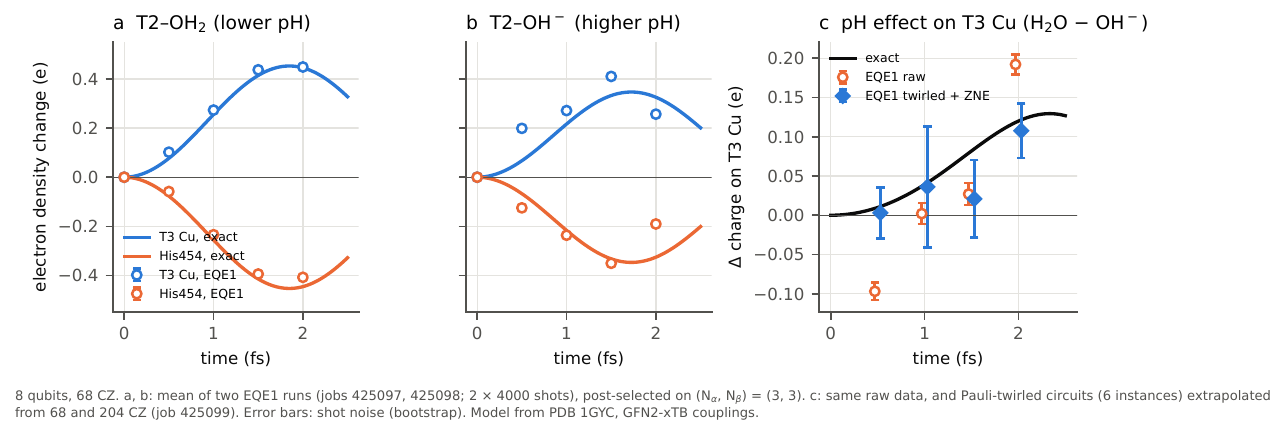}
\caption{Site populations of the four-site junction on EQE1 and the pH effect on T3A.}
\end{figure}
\begin{figure}[h]
\includegraphics[width=\textwidth]{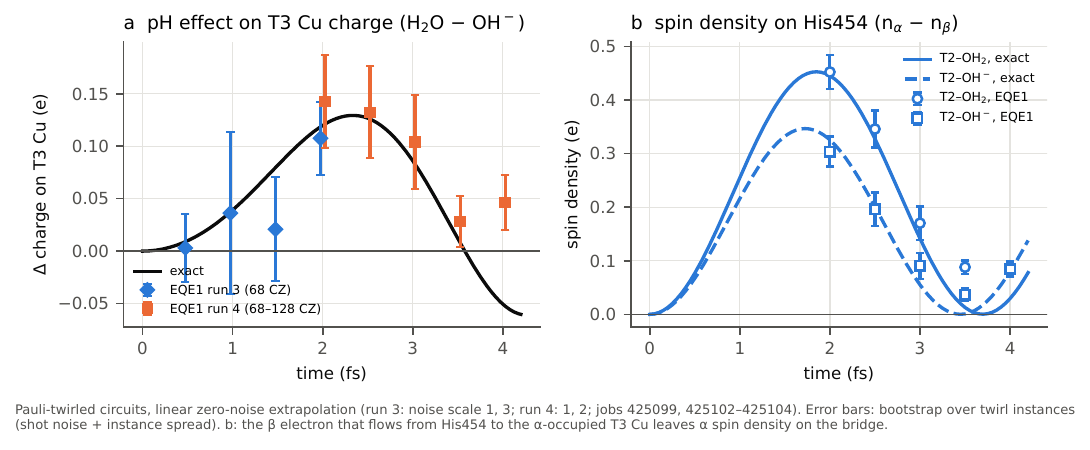}
\caption{pH effect on the T3 charge and the spin density on His454 with zero-noise extrapolation.\cite{Temme2017}}
\end{figure}
\begin{figure}[h]
\includegraphics[width=\textwidth]{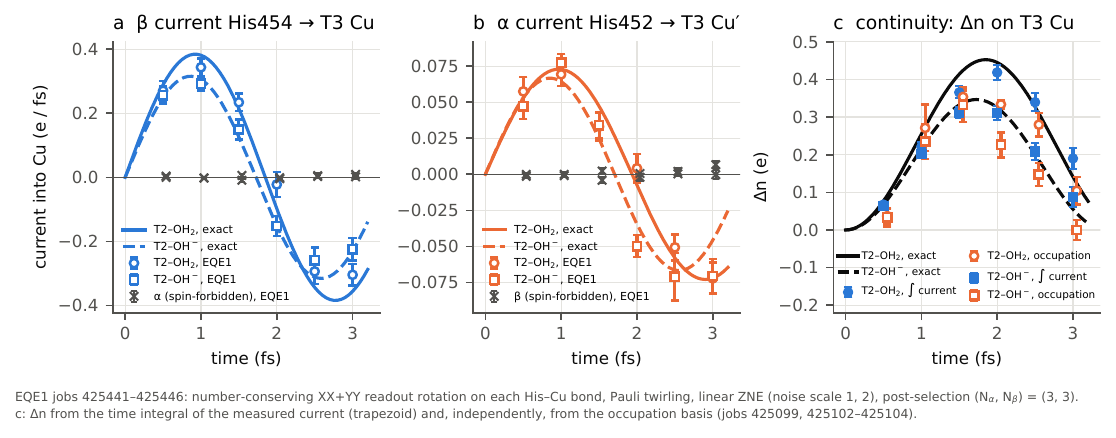}
\caption{Spin-resolved bond currents and the continuity check. The $\alpha$ His454--T3A current vanishes because the
$\alpha$ electrons on His454 and T3A are both present and cannot move (the same reason makes the measured dynamics
effectively single-particle, main text).}
\end{figure}
\begin{figure}[h]
\includegraphics[width=\textwidth]{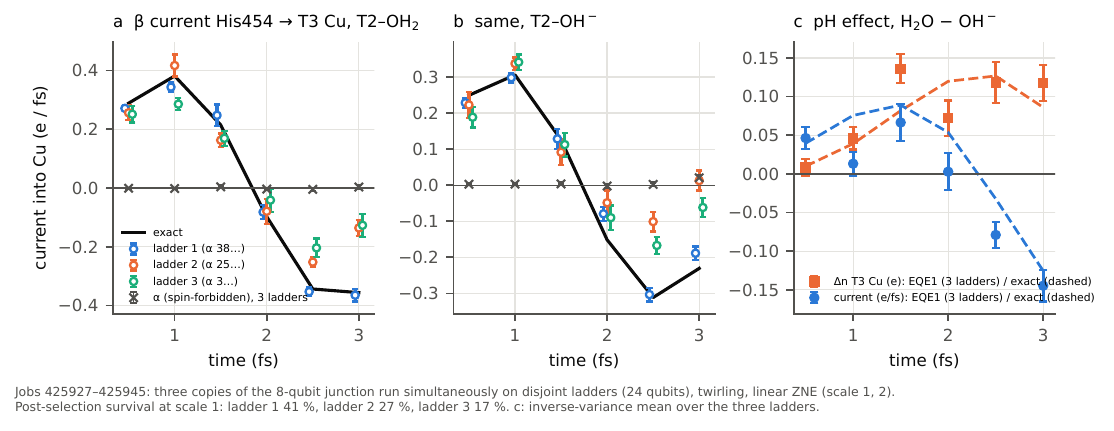}
\caption{Three copies of the junction executed simultaneously on disjoint ladders (24 qubits).}
\end{figure}
\begin{figure}[h]
\includegraphics[width=\textwidth]{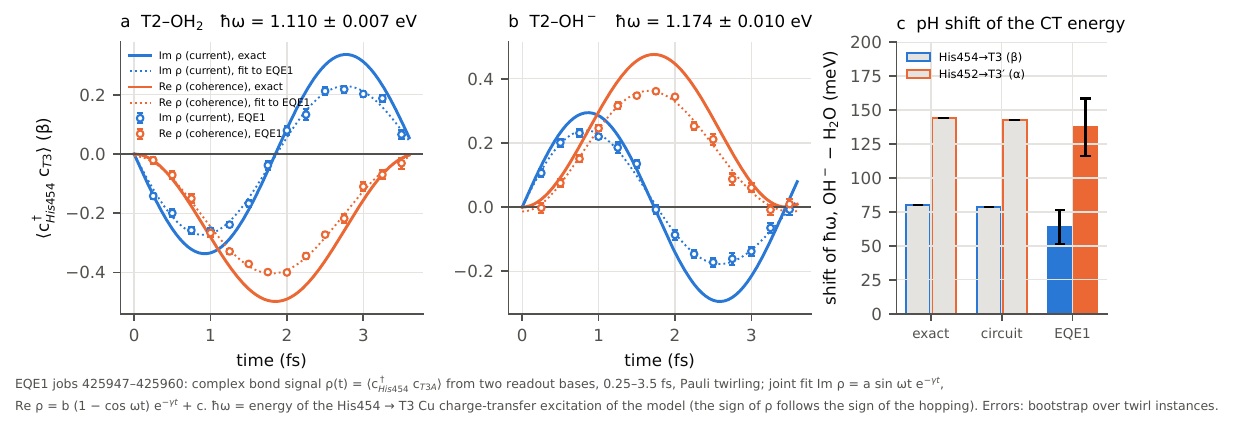}
\caption{First CT-frequency runs (0.25--3.5~fs).}
\end{figure}
\begin{figure}[h]
\includegraphics[width=\textwidth]{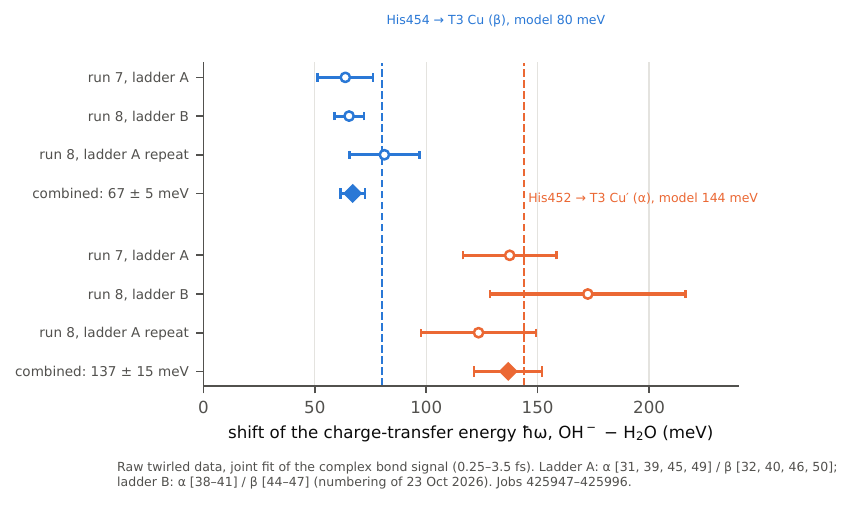}
\caption{Reproducibility of the pH shift of the CT energy across runs and ladders.}
\end{figure}
\begin{figure}[h]
\includegraphics[width=\textwidth]{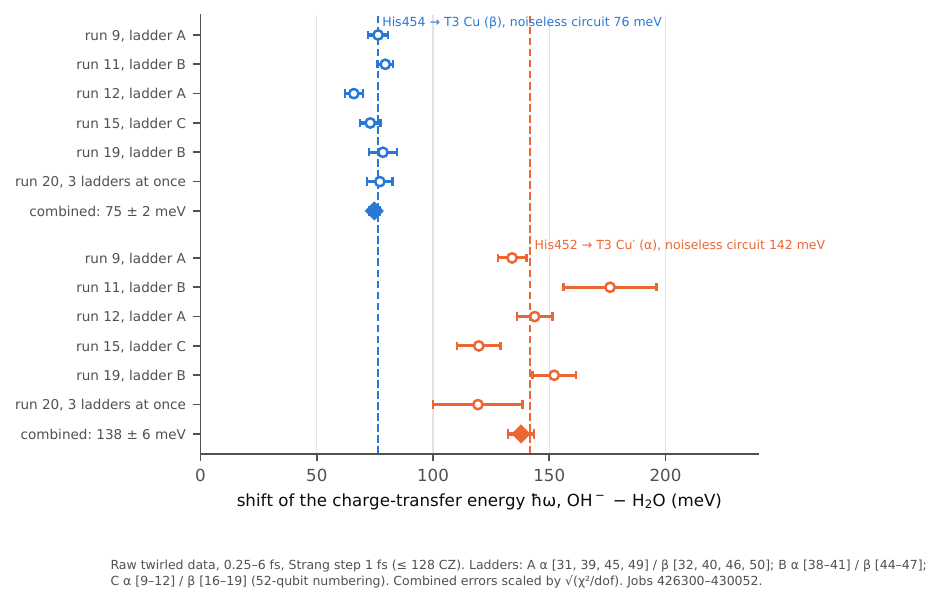}
\caption{CT shift on three ladders, 0.25--6~fs.}
\end{figure}

\section{Practical estimate for pH and chloride}
\begin{figure}[h]
\includegraphics[width=\textwidth]{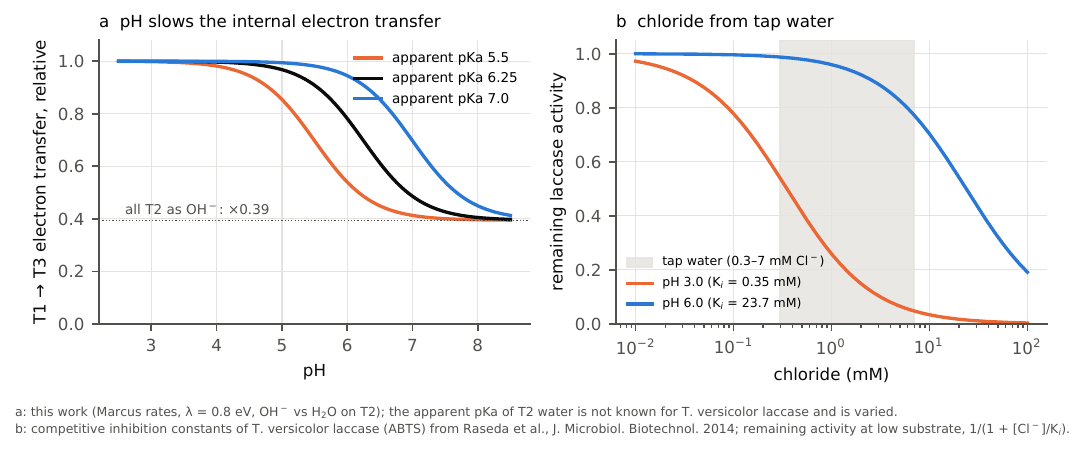}
\caption{Estimate of the internal electron-transfer slow-down with pH (Marcus rate ratio from the junction model and
an assumed apparent p$K_a$) and of chloride inhibition with literature $K_i$ values.\cite{Raseda2014} An estimate,
not a measurement.}
\end{figure}

\section{Attempts beyond the reach of the hardware}
\textbf{Correlated bridge model.} A five-site model including the bridging hydroxide of the T3 pair (10 qubits,
fermionic swaps, 80/152/228~CZ for 1--3 Trotter steps) predicts a pH-dependent redirection of charge to T3B. On two
ladders the post-selection survival was 4--12\% and the effect was seen only as a hint
($+0.017 \pm 0.007$~e against $+0.05$~e for the noiseless circuits).

\textbf{Vibronic model.} An electron coupled to an effective vibrational mode ($\lambda = 0.43$~eV,
$\hbar\omega = 79$~meV; 4 qubits, $\approx$70~CZ) was not resolvable: gate noise heated the mode qubits to
populations of 0.3--0.5 within 3~fs, above the maximal vibronic excitation of 0.2.

\textbf{T1 $\to$ TNC step.} In the one-orbital-per-site description the Cys453--His454 coupling is 5--8~meV in all
models (period $\approx$90~fs, far beyond the $\approx$6~fs window of the hardware signal), because the S~3p lone pair
and the imidazole $\pi$ orbital are nearly orthogonal across the peptide link. The driving force
$\varepsilon(\mathrm{T1}) - \varepsilon(\mathrm{T3A})$ stays positive for WT and Y116E in all dielectric models;
Y116E lowers it by 26--61~meV.

\textbf{Mitigation.} Linear zero-noise extrapolation\cite{Temme2017,Cai2023} improved early junction runs only
slightly and introduced a systematic bias of 0.03--0.05; below about 30\% post-selection survival it no longer
restored the signal.

\section{Real Glu116 and protonation}
\begin{table}[h]
\caption{OH$^-$ penalty on the His454 channel (meV) for the real Glu116 conformers with and without Asp456
(GFN2-xTB, $\varepsilon \approx 4$).}
\begin{tabular}{lrrrrr}
\toprule
 & a & b & c & d & e \\ \midrule
Glu116$^-$ (no Asp456) & 58 & 56 & 54 & 62 & 68 \\
Glu116$^-$ + Asp456$^-$ & 52 & 50 & 49 & 54 & 58 \\
Glu116$^-$ + Asp456-H & 58 & 56 & 54 & 62 & 67 \\
p$K_a$ Glu116 (PROPKA) & 5.6 & 6.4 & 5.0 & 8.5 & 8.9 \\
p$K_a$ Asp456 (PROPKA; WT 7.2) & 8.5 & 8.3 & 9.2 & 6.8 & 6.7 \\
\bottomrule
\end{tabular}
\\[2pt] WT: 87 (no Asp456), 74 (Asp456$^-$), 83 (Asp456-H).
\end{table}

\section{Partial relaxation}
\begin{table}[h]
\caption{OH$^-$ penalty on the His454 channel (meV, GFN2-xTB, $\varepsilon \approx 4$) before and after relaxing the
T2 ligand, the residue-116 group and all hydrogens (other heavy atoms fixed).}
\begin{tabular}{lrr}
\toprule
model & unrelaxed & relaxed \\ \midrule
WT & 87 & 200 \\
WT + Asp456$^-$ & 74 & 201 \\
Y116E probe & 20 & 176 \\
Glu116 c + Asp456-H & 54 & 177 \\ \midrule
reduction, probe & 66 & 24 \\
reduction, real mutant & 20 & 24 \\
\bottomrule
\end{tabular}
\\[2pt] Relaxed hydroxide: Cu(T2)--O 1.93~\AA, O--Cu(T3A) 4.0~\AA{} (crystal geometry 2.35 and 5.9~\AA). The relaxation of
the Y116E-probe OH$^-$ model stopped at the 150-step limit.
\end{table}

\section{DFT checks of the tight-binding parameters}
\textbf{Full clusters (not completed).} A DFT check on reduced clusters (no T1 site and no Cys453; 96--104 atoms,
3 Cu; UPBE, density fitting, gas phase) could not be converged with the available resources: DIIS stalled at
$|g| \approx 10^{-3}$, and a second-order restart converged towards a different, higher-energy open-shell solution.

\textbf{Fragment test of the mechanism.} A fragment cut from the crystal-geometry clusters contains T3A Cu with all its
donor ligands (His454 with its backbone, the other ligand molecules and the bridging hydroxide) and the T2 water
(33 atoms, charge +1, doublet); the perturbation is the residue-116 group of the probe or of Glu116 conformer c
(charge $-1$). With the fragment-orbital procedure of the main text, the shift of the gap
$\varepsilon(\mathrm{His454}) - \varepsilon(\mathrm{T3A})$ caused by the charged group is $-295$~meV (GFN2-xTB) and
$-1554$~meV (UB3LYP; def2-SVP on Cu, 6-31G elsewhere) for Glu116, and $+48$ and $-418$~meV for the probe. For the real
glutamate both methods agree on the direction of the mechanism, a lower His--Cu gap, but not on its size; for the probe
they disagree in sign. A gas-phase fragment without screening, in which an uncompensated anion shifts the Fermi level
and thereby the choice of the Cu orbital, is too crude for a quantitative comparison, and we do not use it as one.

\section{Software}
tblite 0.8.0 (GFN2-xTB),\cite{Bannwarth2019} ffsim 0.0.85, Qiskit 2.6.1,\cite{Qiskit2024} gemmi 0.7.5, NumPy 2.5.3,
SciPy 1.18.1, PROPKA 3.5.1,\cite{Olsson2011} PySCF 2.14.0.

\fi

\end{document}